\documentclass[a4paper]{article}

\usepackage{icrc2013}
  \usepackage[english]{babel}
\usepackage{multicol}
  
\def\lat{{\em Fermi}-LAT}

\title{High confidence AGN candidates among unidentified Fermi-LAT sources via statistical classification}

\shorttitle{High confidence AGN in 2FGL unidentified sources}

\authors{
M.~Doert$^{1,3}$,
M.~Errando$^{2}$
}

\afiliations{
$^1$ Fakult\"{a}t Physik, Technische Universit\"{a}t Dortmund, 44221 Dortmund, Germany \\
$^2$ Department of Physics and Astronomy, Barnard College, Columbia University, NY 10027, USA\\
\scriptsize{
$^3$ now at Department of Physics, Columbia University, New York, NY 10027, USA
}
}
\email{mdoert@nevis.columbia.edu, errando@astro.columbia.edu}

\abstract{The second \textit{Fermi}-LAT source catalog (2FGL) is the deepest survey of the gamma-ray sky ever compiled, containing 1873 sources that constitute a very complete sample 
down to an energy flux of $\sim 10^{-11}$ erg cm$^{-2}$ s$^{-1}$. While counterparts at lower frequencies have been found for a large fraction of 2FGL sources, active galactic nuclei (AGN) being the most numerous class, 576 gamma-ray sources remain unassociated. 
In these proceedings, we describe a statistical algorithm that finds candidate AGNs in the sample of unassociated 2FGL sources by identifying targets whose gamma-ray properties resemble those of known AGNs.
Using two complementary learning algorithms and intersecting the high-probability classifications from both methods, we increase the confidence of the method and reduce the false-association rate to 11\%.
Our study finds a high-confidence sample of 231 AGN candidates among the population of 2FGL unassociated sources.
Selecting sources out of this sample for follow-up observations or studies of archival data will substantially increase the probability to identify possible counterparts at other wavelengths.}


\begin{document}
\maketitle

\section{Introduction}
The second \lat\  source catalog (2FGL) characterizes 1873 gamma-ray sources detected in the energy 
range of 0.1 to 100\,GeV \cite{2012ApJS..199...31N}. The catalog covers the whole sky with little observational bias, although 
the sensitivity is not uniform mainly due to the intensity of the diffuse Galactic gamma-ray 
emission. A total of 127 sources from the 2FGL catalog are firmly identified through simultaneous 
variability (periodic or episodic) or common morphology with their multiwavelength counterparts. An additional 1170 sources are reliably associated with counterparts from a-priori selected  
catalogs of candidate gamma-ray emitting source classes. The remaining 576 sources for which no counterpart was identified are left unassociated. \\
There are fourteen classes of gamma-ray sources represented in the 2FGL catalog with at least one source count. A complete list of source types can be found in Table~\ref{tab:classes}. 
The most numerous class are active galactic nuclei (AGN), representing 60\% of the catalog. Gamma-ray 
emitting pulsars (4.4\%), pulsar wind nebulae and supernova remnants (adding up to 3.8\%) are other well-represented source classes. The rest of the catalog is distributed in 
unassociated sources (31\%), and sources belonging to source classes with small number counts.

\begin{table} [!thh]
\centering
	\begin{tabular}{cccc}
		\hline
		 2FGL& Description &\# id.& Label \\
			class & &+ \#assoc.&\\
		\hline
		bzb & BL Lac-type blazar& 436 & AGN \\
		bzq & FSRQ-type blazar & 370 & AGN \\
		agu & AGN of uncertain type & 257 & AGN \\
		agn & Non-blazar AGN & 11 & AGN \\
		rdg & Radio galaxy & 12 & AGN \\
		sey & Seyfert galaxy & 6 & AGN \\
		psr & Pulsar& 108 & non-AGN \\
		glc & Globular cluster& 11 & non-AGN \\
		snr & Supernova remnant & 10 & non-AGN \\
		pwn & Pulsar wind nebula& 3 &  non-AGN \\
		spp & SNR / PWN & 58 &  non-AGN \\
		hmb & High-mass binary& 4 & non-AGN \\
		nov & Nova & 1 & non-AGN \\
		gal & Normal galaxy & 6 & non-AGN \\
		sbg & Starburst galaxy& 4 & non-AGN \\
		 & Unassociated sources& 576 & \\
		\hline
\end{tabular}
\caption{List of source classes in the 2FGL catalog.}
\label{tab:classes}
\end{table}
\noindent
With this work, we aim to identify unassociated sources in the 2FGL catalog whose gamma-ray 
properties are similar to those of gamma-ray emitting AGNs. To do that, we train two different 
classification algorithms on the gamma-ray properties of the known AGNs in the 2FGL catalog and apply 
them to the population of unassociated sources.

\section{Source classes in the 2FGL catalog}
The main goal for this work is the identification of high-confidence AGN candidates. We approach the classification of unassociated 2FGL 
sources as a two-class problem, where each source can either be labeled as ``AGN" or ``non-AGN". 
Following this approach, we assigned one of these labels to each 
of the fourteen source classes represented in the 2FGL catalog, as shown in Table~\ref{tab:classes}.
The most abundant classes which contribute to the AGN sample are blazars (both BL Lac and FSRQs), as well as AGN of uncertain type. Most of the ``non-AGN" sources 
are pulsars, pulsar wind nebulae
and supernova remnants. 
The total number of identified and associated sources is 1297,  
out of which we label 1092 elements as ``AGN" and 205 as ``non-AGN". The unassociated sample comprises 576 sources. 
 After the classification, each of these sources will be labelled as either ``AGN", ``non-AGN" or ``unclassified". The latter applies to sources for which the confidence for a correct classification is below a certain 
threshold which we specify during the optimization. This way, an optimal purity
of the sample of sources tentatively labeled 
as AGN is achieved.
 \begin{figure*}[tdp]
 \center
 \includegraphics[width=0.99\textwidth]{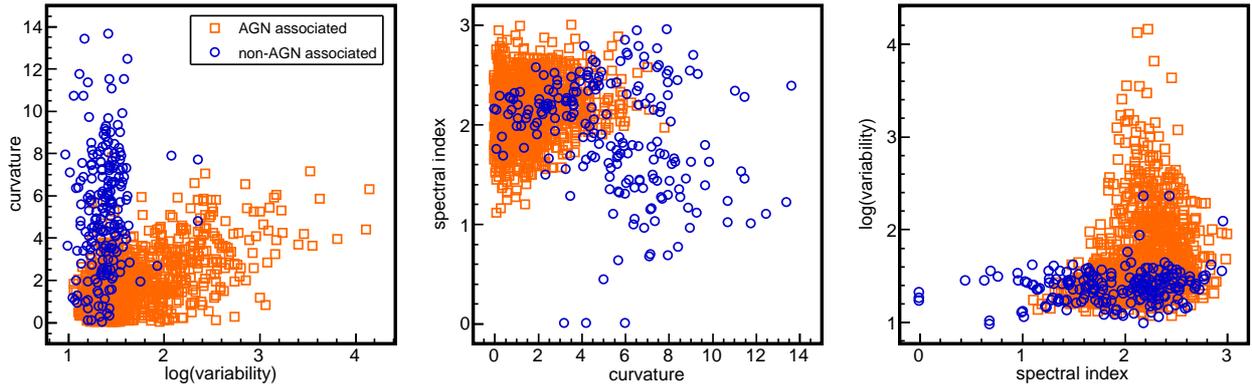} 
 \caption{Scatter plots showing the spectral curvature, flux variability, and spectral index for 2FGL sources associated with AGN and non-AGN sources. The distributions show that the gamma-ray properties of AGN differ from those of the other source classes. \label{fig:2fgl_par}}
 \end{figure*}

\section{Selection of attributes}
\begin{figure*}[htdp]
 \center
 \includegraphics[width=0.50\columnwidth]{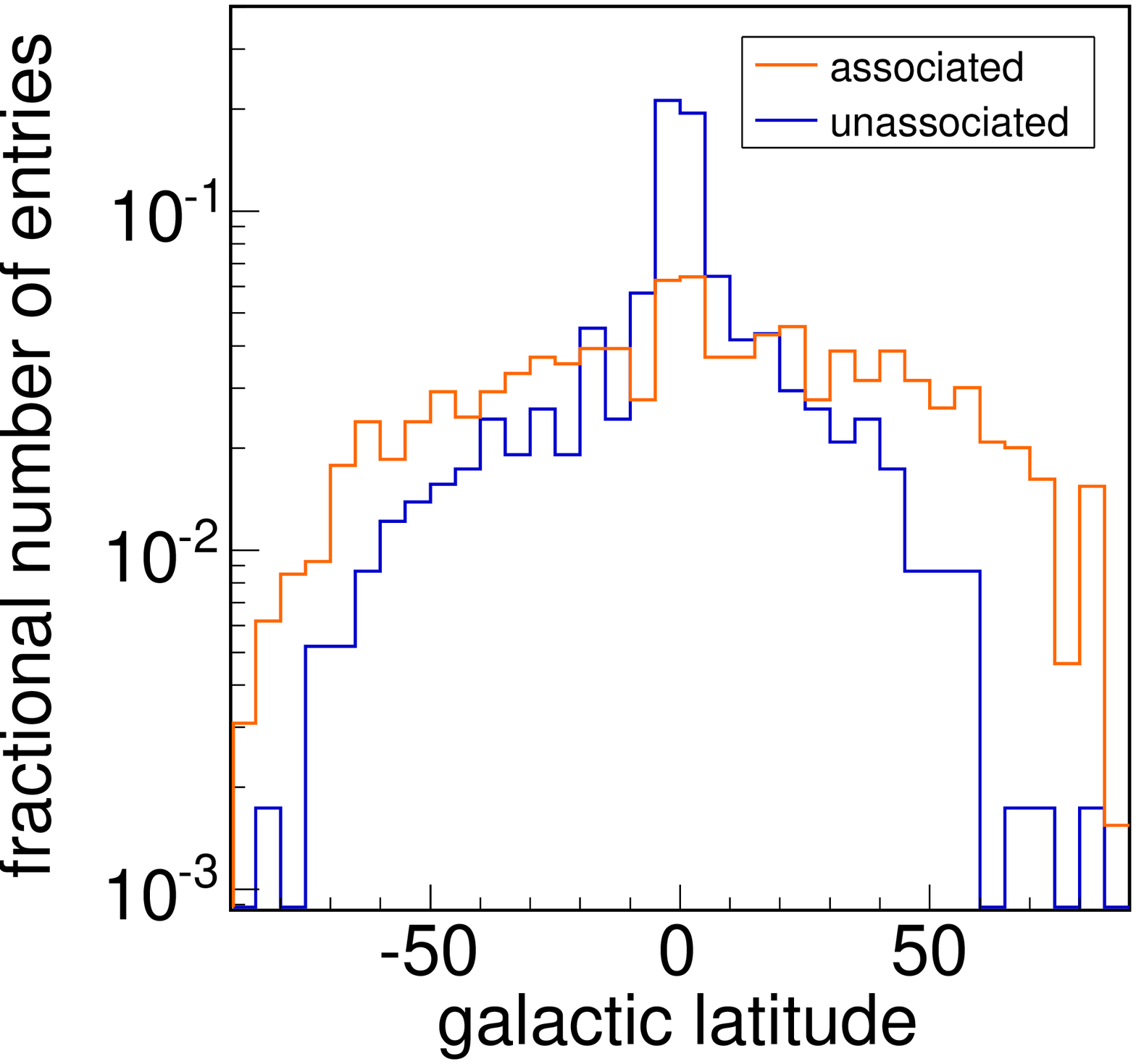} 
 \includegraphics[width=0.50\columnwidth]{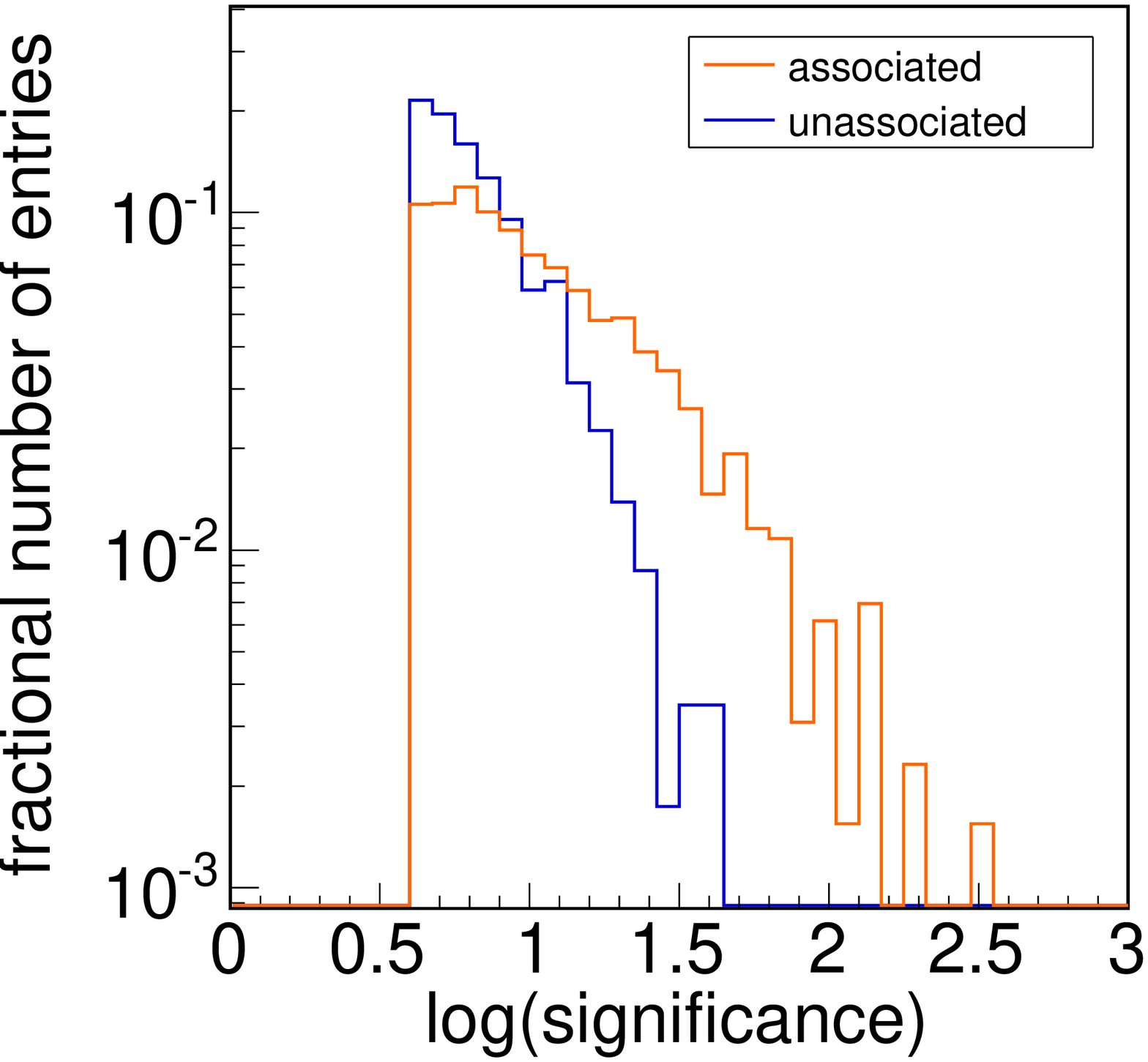}
 \includegraphics[width=0.50\columnwidth]{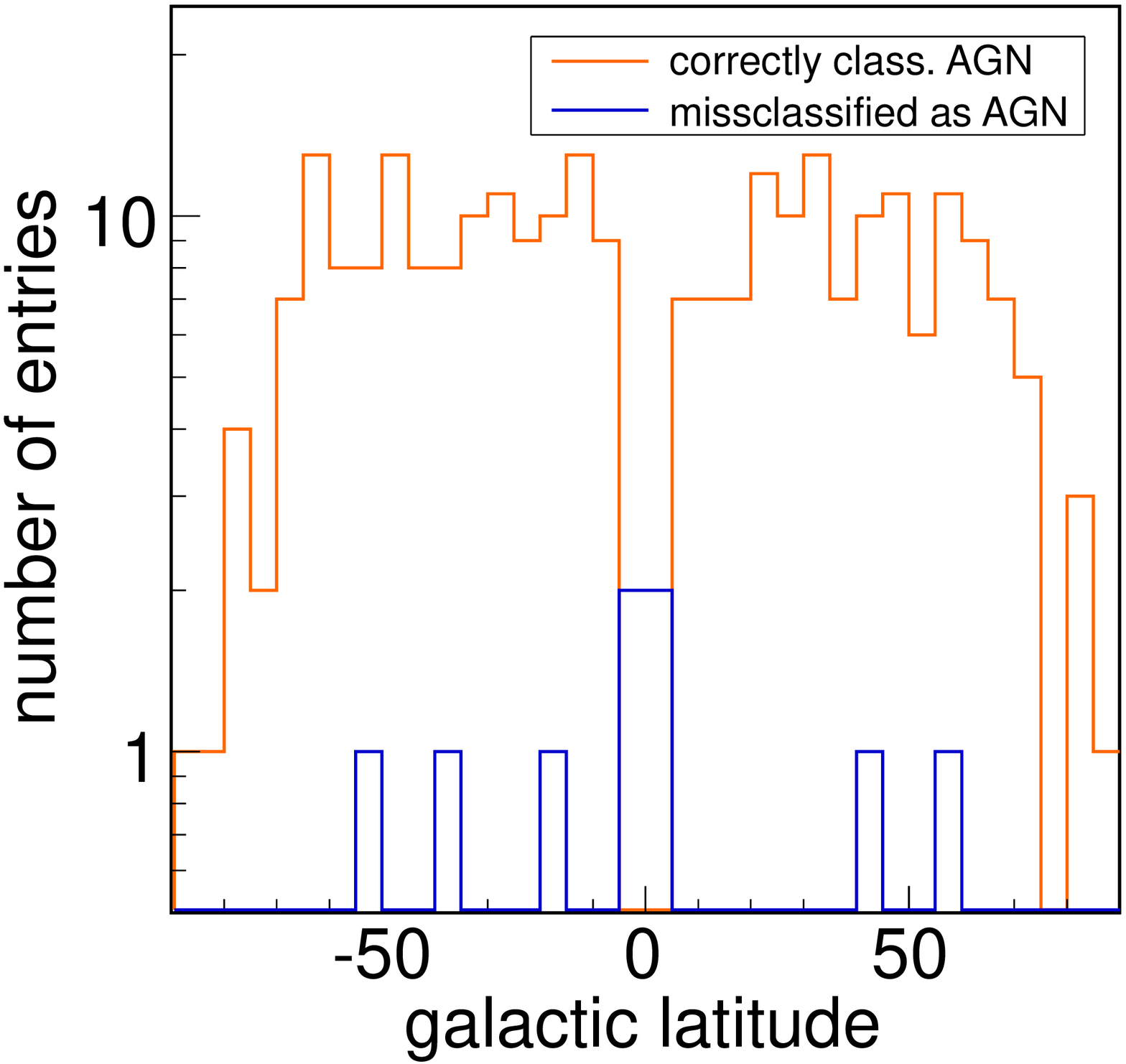} 
 \includegraphics[width=0.50\columnwidth]{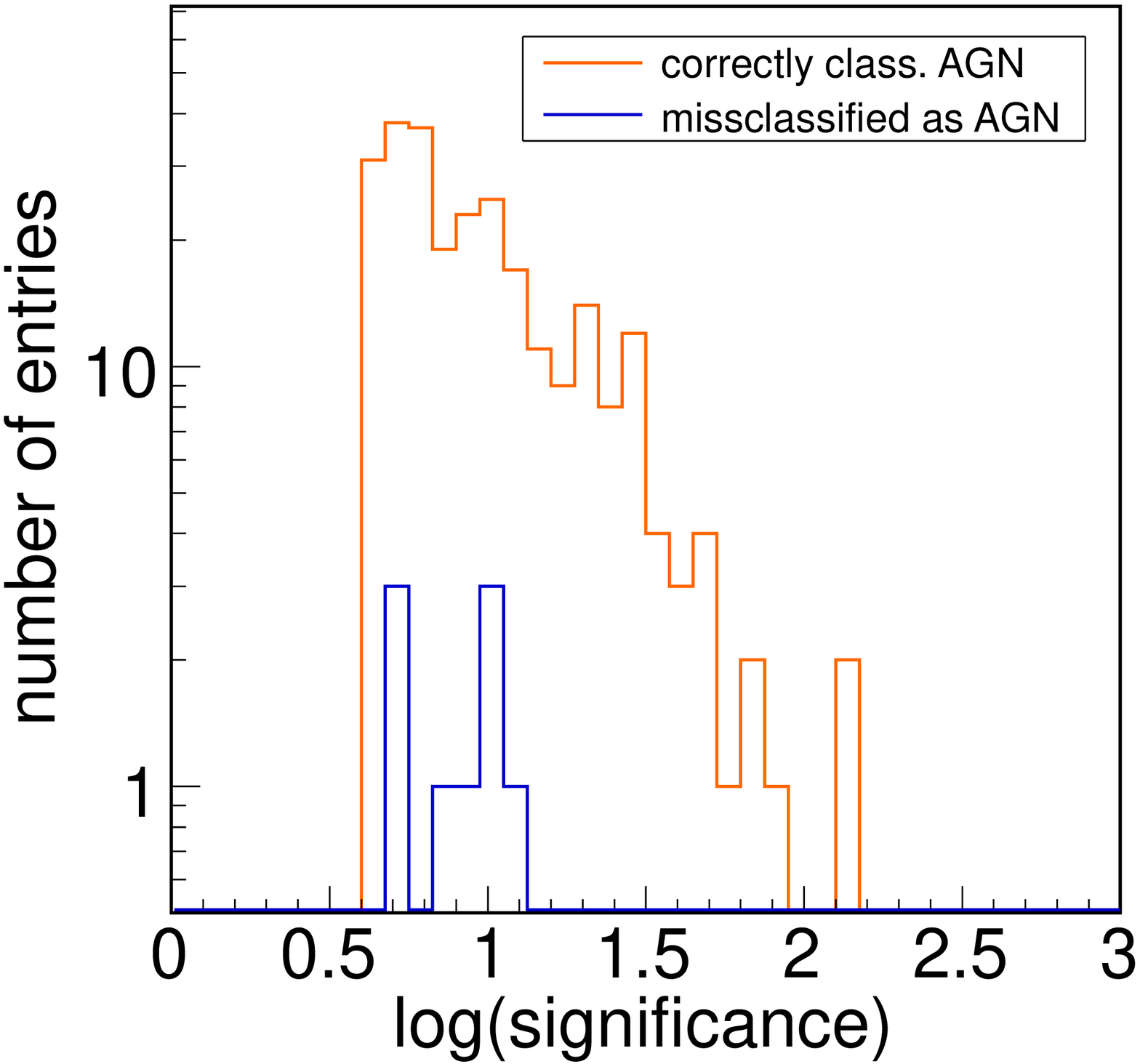} 
 \caption{ Latitude (\emph{left}) and significance (\emph{middle-left}) distributions of the 2FGL sources for the associated and unassociated sample. 
On (\emph{middle-right}) and (\emph{right}), same distributions for the sources in the test sample are split between sources correctly classified as AGN and misclassified as AGN. The distributions show that the efficiency of the classification algorithm drops dramatically for sources at low galactic latitudes and low significance. 
\label{fig:reweighting}}
 \end{figure*}

In the 2FGL catalog, the gamma-ray properties measured by the \textit{Fermi}-LAT are reported for every source. Among these attributes are the \textit{flux values} $F$ for five energy bands with boundaries at 0.1, 0.3, 1, 3, 10 and 100\,GeV, 
the spectral index obtained from a power law fit to the energy spectrum, parameters quantifying the flux \textit{variability} and spectral \textit{curvature}, and the \textit{significance} associated to the detection of each source.
Figure~\ref{fig:2fgl_par} shows distributions of some parameters directly extracted from the 2FGL catalog for sources in the associated sample. Here the AGNs show distinct properties compared to other source classes.
To train our learning algorithms, we explored the set of attributes present in the 2FGL catalog as well as physically meaningful combinations of those, many of them already introduced in \cite{2012ApJ...753...83A}. The best separation power between the populations of ``AGN'' and ``non-AGN'' sources in the catalog was found when using the following attributes: 
\begin{itemize}
\begin{multicols}{2}
\setlength{\itemsep}{3pt}
\item $HR_{12}$
\setlength{\itemsep}{3pt}
\item $HR_{23}$
\setlength{\itemsep}{3pt}
\item $HR_{34}$
\setlength{\itemsep}{3pt}
\item $HR_{45}$
\setlength{\itemsep}{3pt}
\item \textit{hardness slope} 
\setlength{\itemsep}{3pt}
\item \textit{normalized variability} 
\setlength{\itemsep}{3pt}
\item \textit{spectral index}
\vspace{2cm}
\end{multicols}
\end{itemize}
where $HR_{ij}$ describes the hardness ratio between the energy fluxes measured in two contiguous spectral bands:
\begin{equation}
HR_{ij}=\frac{F_i E_i - F_j E_j}{F_i E_i + F_j E_j}
\end{equation}
where $F_i$ and $E_i$ are respectively the flux and mean energy in the $i$-th spectral energy band, with $i=1$ being the lowest spectral band reported in the 2FGL catalog. A \textit{hardness slope} parameter was also defined as 
\begin{equation}
hardness\ slope=HR_{23} - HR_{34}
\end{equation}
which presents a powerful handle to separate possible AGN candidates from pulsar-like sources, as pulsars generally show a spectral cut-off around these energies. 
Additionally, we use a \textit{normalized variability}  defined as 
\begin{equation}
normalized\ variability=\frac{variability}{significance}
\end{equation}
We do not use any variable that is directly related to the overall flux of the sources detected by \lat\ as associated sources have on average higher fluxes and detection significances than the sources in the unassociated sample (see Figure~\ref{fig:reweighting}, middle-left panel). 
To avoid an influence of the different ranges of the individual parameters on the classification, we renormalized all attribute distributions to the range between 0 and 1.

\section{Analysis tools and Methods}
We perform the complete classification process within the data mining framework Rapid Miner. This is an open-source software originally developed at Technische Universit\"{a}t Dortmund under the name ``YALE'' and now maintained and distributed by Rapid-i \cite{Mierswa2006}. It offers attribute selection, combination and filtering tools, as well as a variety of built-in classification methods. \\
Before starting the learning process, we split the associated 2FGL sources (1297 elements) into a training sample (70\% of the sources) and a test sample (30\%) using stratified sampling. The training sample is used to train the learning algorithms and optimize their performance, while the test sample is set aside and only used after the algorithms have been trained and optimized to evaluate their performance.\\
We investigated a variety of supervised statistical classification methods. 
To achieve a good estimate of the suitability of each method, we performed a coarse optimization and assessed the performance using ten-fold cross-validation on the training sample. Here, the sample is iteratively trained on 90\% of the training sample and tested on the remaining 10\%, and this process is repeated 10 times until the entire training sample has been tested.\\
\noindent
Based on their robustness and the obtained performance, we chose the \textit{random forest} method (RF) and \textit{neural networks} (NN) as the two algorithms to use in our study. The RF is a very powerful classification algorithm based on the construction of a number of decision trees, where the attributes used for separation in each node are randomly selected \cite{Breiman:2001vj}. 
The NN is a learning method which is based on different layers of interconnected nodes, so-called \textit{neurons}, and has been developed as an artificial replication of the human nervous system \cite{Bertsekas:1995jl}. 

\section{Optimization of the algorithms}
\begin{figure*}[tbdp]
 \center
 \includegraphics[width=0.7\columnwidth]{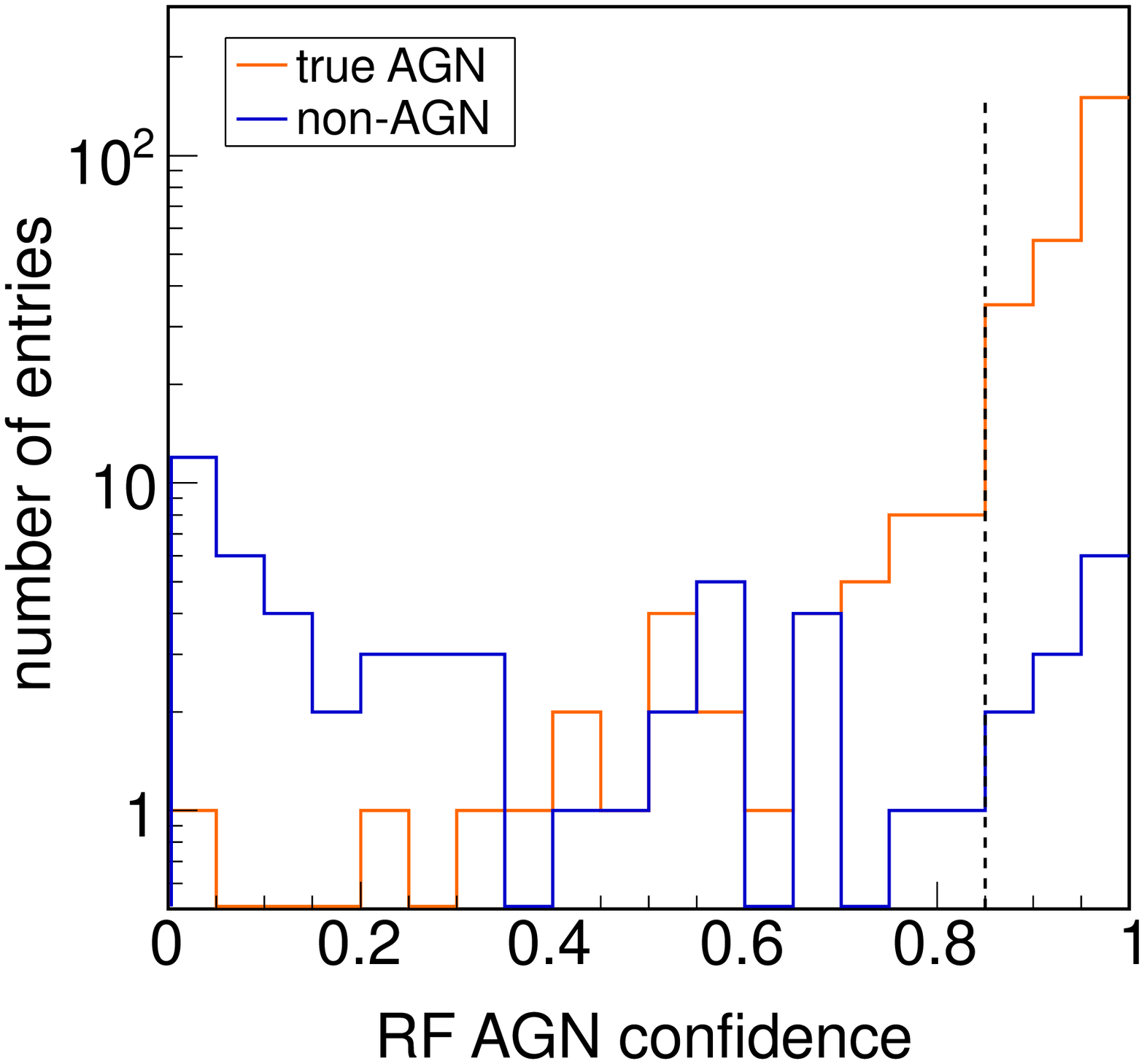} 
 \hspace{2cm}
 \includegraphics[width=0.7\columnwidth]{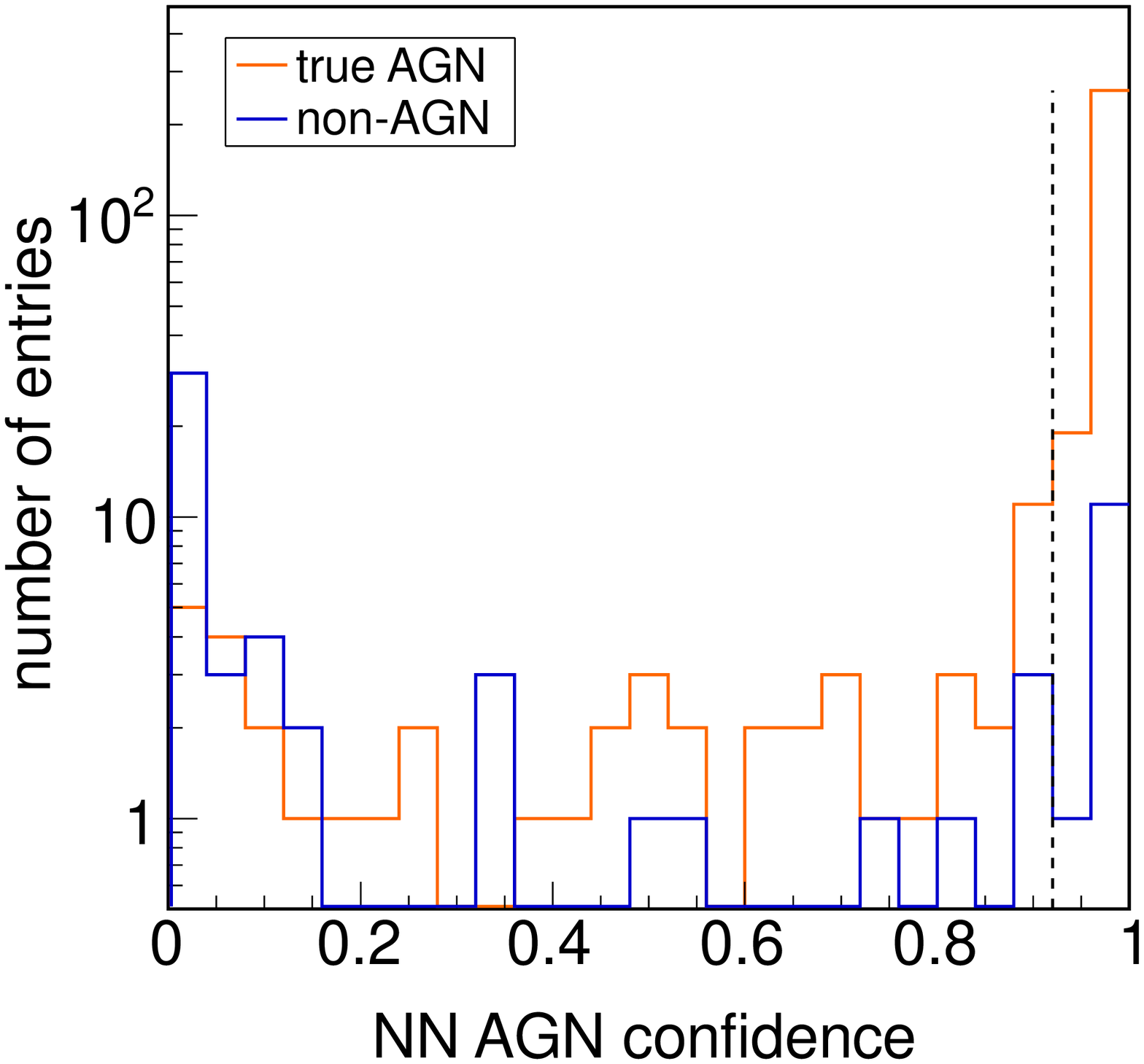} 
 \caption{ Distribution of the confidence of an AGN classification for the sources in our test sample, split between AGN and non-AGN. The distributions are shown for random forest (\emph{left}) and neural networks (\emph{right}). The distributions for AGN peak towards large confidence values, and the vertical dashed line indicates the confidence cut we used for each algorithm.\label{fig:confidence}}
\end{figure*}

We optimized the two selected methods further on the training sample using cross-validation, evaluating the achieved performance in terms of the 
fraction of sources incorrectly classified as AGNs and 
the fraction of true AGNs which are correctly classified as such. 
The two classification methods can be optimized by tuning some key parameters. For RF we used a $number\ of\ trees =100$ and the $depth\ of\ trees =10$, while for NN we selected a $number\ of\ cycles=1000$, $learning\ rate=0.2$, and $momentum=0.1$.\\
\noindent
Each method provides a confidence value for each classified item which gives the probability for this item to be correctly labeled. 
We introduced additional thresholds for both confidence values of sources assigned as ``AGN'' and adjusted them such that we achieved a fraction of misclassified sources in the final sample of $\sim 10\%$. 
The distributions of confidence values for true AGN and non-AGN in the test sample are shown in Fig.~\ref{fig:confidence}, together with the chosen thresholds for each classification method.

\noindent
The final classification algorithm consists in assigning the ``AGN" label only to sources that have been classified as ``AGN" by both classification algorithms independently (NN AND RF) and passed the confidence thresholds which we defined for each of the methods. 
Out of all tested methods, the combination of RF and NN methods was found to have the smallest overlap of the populations of misclassified sources, thus reducing the number of wrongly classified sources. 
A combination of three or more methods did not deliver a considerable gain in performance compared to the loss of correctly classified sources in the final sample. 

\section{Estimation of the performance}
We assess the performance of the combined algorithms and thresholds in terms of the \textit{false-association rate}, i.e.~the fraction of sources incorrectly classified as AGNs, and the \textit{recall}, which is defined as the fraction of true AGNs which are correctly classified and appear in the final sample. 

\noindent
The populations of associated and unassociated 2FGL sources have different distributions in significance and galactic latitude (see Fig.~\ref{fig:reweighting} left panels). Weaker sources with lower significance have larger location errors and are less likely to be present in counterpart catalogs or cannot be identified unequivocally, and are therefore more abundant in the unassociated sample. The same happens for sources near the galactic plane, where the bright galactic diffuse gamma-ray background makes the location of gamma-ray sources more uncertain, and some counterpart catalogs are incomplete due to the foreground and extinction on the plane at other wavelengths.

\noindent
Low significance and low latitude sources also pose a challenge to the classification methods used in this study. They offer less firm information about, e.g.,~the spectral shape or the variability, and they are influenced by the higher levels of diffuse emission. 
During the optimization of the algorithms, we saw that weak and low-latitude sources were much more likely to be misclassified, as shown in the right panels of Figure~\ref{fig:reweighting}. Therefore, testing the performance of our classification method on the test sample would lead to an over-optimistic estimation of the false association rate, as the sample of unassociated sources contains a larger fraction of low-significance and low-latitude sources that our algorithms are more likely to misclassify. 
To get a realistic estimation of the false-association rate, we assign weights to each source in our training and test samples according to their significance and galactic latitude, so that the weighted distributions match that of the population of unassociated sources. Then, the false-association rate is estimated by dividing the sum of the weights of the sources misclassified as AGNs by the total sum of weights. 

\noindent
Table~\ref{tab:test} shows the performance of each individual algorithm and of the combined algorithm which requires for each source to be labeled as ``AGN'' by both methods (NN AND RF) in order to be classified as such. The combined algorithm is expected to recognize 80\% of the AGNs present in the sample of unassociated 
sources, while having an estimated contamination of sources incorrectly labeled as AGNs of 11\%. 
\begin{table} [!thh]
\centering
	\begin{tabular}{lccc}
		\hline
		 & NN & RF & NN AND RF\\
		\hline
		recall & 84.7\% & 88.1\% & 79.6\% \\
		false-assoc. rate & 13.0\% & 16.3\% & 11.2\% \\
		\hline
\end{tabular}
\caption{Performance of the neural networks (NN), random forest (RF), and combined 
algorithm (NN AND RF) evaluated on the weighted test sample of classified 2FGL sources. For completely uncorrelated algorithms, the expectation for the combined performance would be 74.6\% in recall and a false-association rate of 10.2\%.}
\label{tab:test}
\end{table}
\noindent
\section{Results}
\begin{figure}[tdp]
 \center
 \includegraphics[width=0.99\columnwidth]{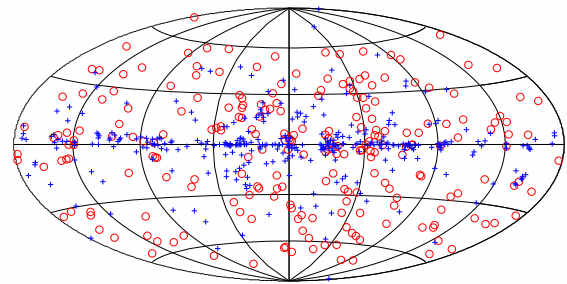} 
 \caption{Sky distribution in galactic coordinates of all unassociated 2FGL sources. Sources labeled as AGN by our classification method are shown as red circles, while blue crosses indicate sources that were not classified as AGN. \label{fig:skymap}}
 \end{figure}

After evaluating the performance on the test sample, we apply the combined classification method 
(NN AND RF) to the sample of unassociated sources. From a total of 576 sources, 231 are 
labeled as AGNs. According to the performance estimated on the test sample, up to 26 of the 
231 tentative classifications are expected to be non-AGNs whose gamma-ray properties are similar to 
those of the known 2FGL AGNs. The sky distribution of the AGN candidates, together with the sources that were not conclusively labeled as candidate AGNs, is shown in Figure~\ref{fig:skymap}. 
\section{Conclusions}
We have used two independent classification algorithms to find objects in the unclassified sample of the 2FGL catalog whose gamma-ray properties resemble those of gamma-ray emitting AGNs, the most numerous source class detected by \lat.
Our work identifies 231 AGN candidates based on their gamma-ray properties, which constitute 40\% of the 2FGL unassociated source population. The final list of sources is available upon request. 

\noindent
By weighting the sample of test sources to have a similar significance and galactic latitude distribution as the unassociated sample, we calculate a realistic false-association rate of 11\%. This fraction is significantly higher  than the $0.05\%$ reported in \cite{2012ApJ...753...83A}, which was estimated on the fraction of wrong associations in the test sample. However, the value in \cite{2012ApJ...753...83A} is likely to be an over-optimistic estimate, since most misclassifications occur for sources with low galactic latitude or small detection significance, which are more abundant in the unassociated source sample than in the test sample. 
In a similar work, a false-association rate of 2.3\% was reported in \cite{2012MNRAS.424L..64M} estimated using cross-validation. That estimate is less sensitive to low-significance and low-latitude sources as the authors excluded the galactic plane from their study. However, cross-validation is known to give optimistic performance estimates as the same population of sources is used for training and testing. \\ 
Studying and identifying the selected AGN candidates  through multi-wavelength studies is likely going to extend the population of gamma-ray emitting AGNs to lower gamma-ray fluxes and therefore lower luminosities, having a potential impact on population studies and the estimates of the contribution of unresolved AGNs to the extragalactic diffuse gamma-ray emission. In addition, our classification method can also help in targeting unassociated AGNs close to the galactic plane, where counterparts are more difficult to identify due to galactic extinction and diffuse foreground emission at low galactic latitudes.

\vspace*{0.3cm}
\noindent
\footnotesize{{\bf Acknowledgments}{\\
We gratefully acknowledge support by DFG (SFB\,823, SFB\,876), DAAD (PPP USA) and HAP. 
ME acknowledges support from NASA grant NNX12AJ30G. We thank Sabrina Einecke, Brian Humensky, Reshmi Mukherjee, Daniel Nieto and Ann-Kristin Overkemping for feedback on the manuscript.
}}

\end{document}